# GENETIC AGENT APPROACH FOR IMPROVING ON-THE-FLY WEB MAP GENERALIZATION

Brahim lejdel[1] and Okba kazar[2]

[1]Computer science departement, University of el-Oued, Algeria.

`lejdel82@yahoo.fr`

[2]Computer science department, University of Biskra, Algeria.

`kazarokba@yahoo.fr`

## ABSTRACT

*The utilization of web mapping becomes increasingly important in the domain of cartography. Users want access to spatial data on the web specific to their needs. For this reason, different approaches were appeared for generating on-the-fly the maps demanded by users, but those not suffice for guide a flexible and efficient process. Thus, new approach must be developed for improving this process according to the user needs. This work focuses on defining a new strategy which improves on-the-fly map generalization process and resolves the spatial conflicts. This approach uses the multiple representation and cartographic generalization. The map generalization process is based on the implementation of multi-agent system where each agent was equipped with a genetic patrimony.*

## 1. INTRODUCTION

The web mapping has known great growth in parallel of the rapid development of the internet. To provide on-the-fly web mapping to the user, the process of on-the-fly map generalization must rely on fast, effective, and powerful methods. An principals challenge of such on-the-fly maps generalization is to offer the user a spatial data in real time and in height quality , it must allow also to solve spatial conflicts that may appear between objects especially due to lack of space on display screens [1].

Many research works have proposed solutions to on-the-fly map generalization, an important work that was done by [1], he proposed a new approach based on the implementation of a multi-agent system for the generation of maps on the fly and the resolution of spatial conflicts. This approach is based on the use of multiple representation and cartographic generalization. Also, to improve the process of on-the-fly map generalization, another approach was proposed in [2] which based on a new concept called SGO (Self-generalizing object). In the same context and for reducing the spatial conflicts in the map, a good method was proposed in [3] which based on the genetic algorithm. The work presented in this paper comes under the research favouring the use of agents were equipped with genetic patrimony; it based on the approach presented in [4]. It uses the genetic agent in order to generate data on arbitrary scales thanks to an on-the-fly map generalization process. The proposed approach exclusively aims to solve the following problems related to on-the-fly web mapping applications:

- How can we adapt the contents of maps to users' needs and improve the time of generalization process of spatial data, which improve in result the transfer time of data map?

- How can we solve spatial conflicts in order to improve the quality of map?

The paper is structured as follows. In the part 2, we present on-the-fly map generalization process and its problems. Part 3 briefly reminds the approaches of on-the-fly map generalization. Part 4 presents the proposed approach for improving on-the-fly web mapping generalization. In part 5, we present the experimentation and some results. Finally, part 6 concludes and cites some perspectives of this work.

## 2. ON-THE-FLY WEB MAP GENERALIZATION

The on-the-fly web map generalization is defined as; the creation in real-time and according to the user's request, of a cartographic product appropriate to its scale and purpose, from a largest-scale database. The main characteristics of on-the-fly web mapping are [1]:

- Required maps must be generated in real-time.

- Generation of a temporary and reduced scale dataset for visualization purposes from the database [5] in order to use the computer's memory efficiently [6].

- A real-time map generation process has to take into account users' preferences and contexts.

- A real-time map generation process must adapt maps' contents to display space and resolution of display media as well as to the contextual use of these maps.

- The scale and theme of the map are not predefined [6].

- There is no way to verify the quality of the final map that will be sent to the user.

The main problems linked to on-the-fly map generalization are the time of delivering the cartographic data and its quality. The generalization process time is a crucial factor to provide a user cartographic data. The waiting time must be compatible with Newell's cognitive band, which is less than 10 seconds [7]. Also, in order to produce maps suited to a user's requests, on-the-fly map generalization must be flexible enough to take into account the level of detail, the kind of the map…etc.

## 3. APPROACHES OF ON-THE-FLY MAP GENERALIZATION

There are three fundamental approaches to providing on the fly map generalization:

### 3.1 Generalization-oriented approaches

This approach based on map generalization which is known to be a complex and time consuming process. In order to accelerate this process, certain authors propose methods based upon pre-computed attributes [1]. Cartographic generalization operators have to be applied to spatial objects on-the-fly. The generalization operators are principally the selection, the simplification and the displacement. The generalization oriented approach is very flexible [2]. However, it is not widely used because of the time it takes to provide requested maps. Furthermore, due to its complexity, generalization process cannot be carried out by simply applying generalization algorithms sequentially without taking into account the objects' spatial neighbourhood.

## 3.2 Representation-oriented approaches

Currently, it is the ideal solution to allow users to get data at the desired level of abstraction compared by the previous approach. This approach proposes to store several pre-defined representations of a given object (usually at different scales) within the same database [2]. The simplest representations are usually obtained from the manual or semi-automatic generalizations of the most detailed representations. However, in terms of personalization, multi-representations are extremely limited because all scales are predefined. Other important problems related to multiple representations are the difficulty to create necessary map scales [2]. All these problems limit the effective use of multi-representations for on-the-fly map generalization.

## 3.3 Hybrid approaches

Hybrid approaches take advantage of the flexibility of generalization oriented approaches and the suitability of representation-oriented approaches to generate maps in real-time by combining their use [1]. Several authors proposed an approach based on this hybrid approach, such as [1], [2] and [6]. Its advantage is that it reduces the effort needed for generalization process and improves the quality of the result because smaller the difference between the initial map scale and the desired one, easier the generalization process. However, to be truly efficient, this method must rely on a database that includes several scales, leading to the typical problems associated with multiple representations [2].

To improve this third approach, it is necessary to develop new methods that minimize, as much as possible, the problems associated with automatic generalization and multiple representations and resolve all the spatial conflicts.

## 4. PROPODED APPRAOCH

This work is based on the approach proposed in [4]. We combine genetic agent, map generalization process and multiple representations approach for improving the delivery time of map and resolving spatial conflicts to increase the quality of result map. This approach aims exclusively to improve the map generalization process. The spatial objects are modelled as agent. This agent is a concept of artificial intelligent. Each agent is equipped with genetic patrimony. Thus, genetic agent has some knowledge of its internal state, and some sensory information concerning environmental context, which permit it to decide what action (or action sequence) executed in order to achieve its goals.

In this context, the role of genetic agent is identifying the best sequence of generalization's algorithms with good parameters that allow to perform the best map generalization process. In order to implement this process, each agent is able to identify and assess its internal constraints, and applies generalisation algorithms to itself in order to satisfy as well as possible these constraints. They are divided into two kinds [8]:

-Internal constraints: The constraints relate an isolated object. In this context, we take account for building agent; constraints of size, granularity, squareness, preservation of shape, etc.

-Relationalships constraints: They are the constraints which involve more than one spatial object, the constraint that prevents symbols from overlapping each others (, e.g the symbol of a road should not overlap with that of a house), or the constraint that requires aligned buildings to remain aligned.

In this section, the genetic agent architecture is presented. We describe also the structure of plan executed by the genetic agent, its different modules and the cycle life of genetic agent.

### 4.1. Genetic Agent architecture

As previously mentioned, this research work based upon the use of hybrid approach which allows to improve the generalization process of spatial data described in [4]. This approach consists to model the geographic objects contained in databases (roads, buildings, etc.) into the decisional entities of the generalization process. In this approach, every geographic object becomes a software agent whose goal is to satisfy its constraints as much as possible. Each geographic agent has three main components:

-Patrimony genetic.
-An optimizer: using the genetic algorithm to find the best solution according to the internal constraints or relationships ones.
-Ability to communicate with neighbouring agents.

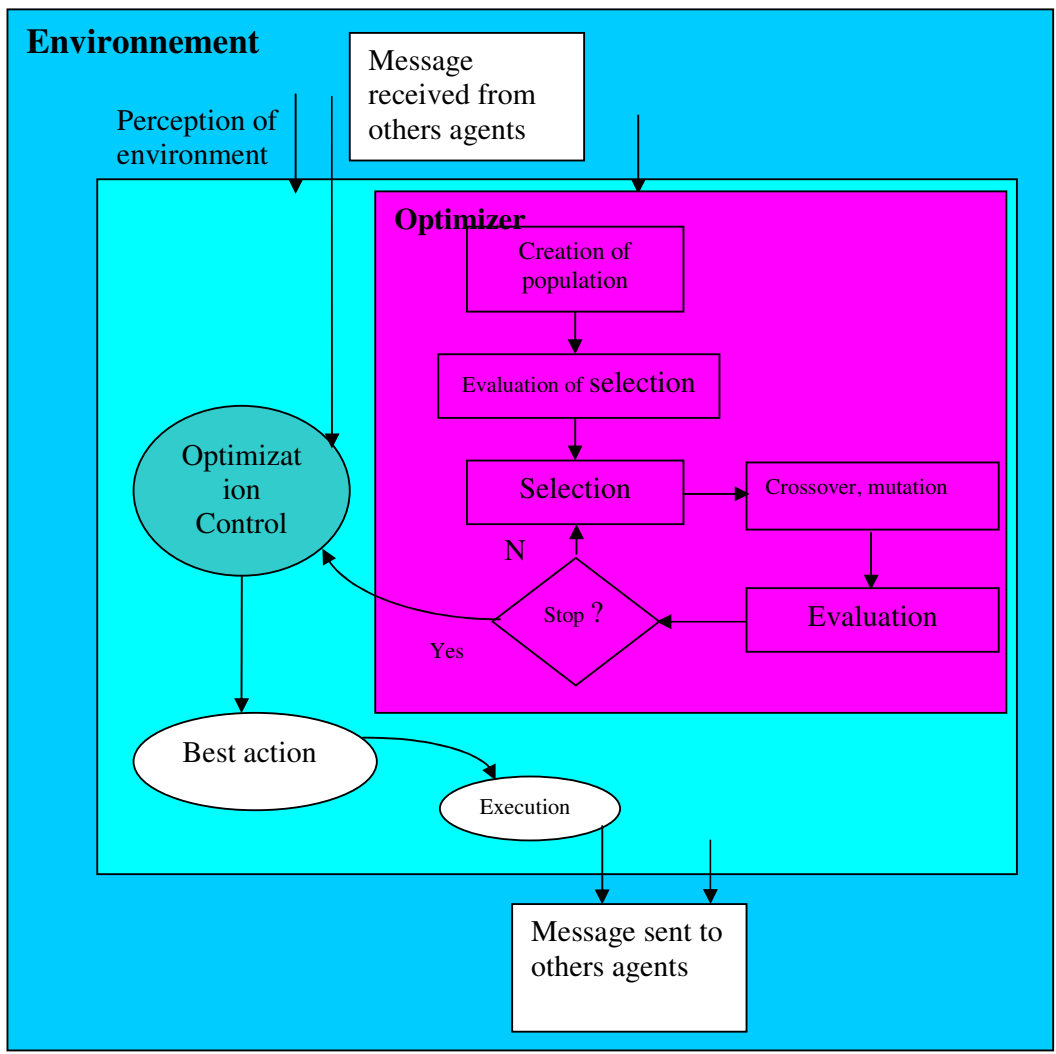

Figure 1. Genetic agent architecture

## 4.2 Plan Structure of genetic agent

The genetic agent is an agent which has a patrimony genetic. In this context, we consider two main kinds of genetic agent; road's genetic agent and building's genetic agent. In the following, we represent the gene of these two kinds:

### 4.2.1 Plan of Road gene

An agent road can be characterized by its identified, and a set of algorithms that can be applied on it, to perform its generalization. These algorithms can take parameters. The coding of gene is made by multiple forms; we use character code for coding the identifiers, the binary form to implement the application or not of such algorithms (0 to say that the algorithm is not applied, 1 for say that the algorithm is applied) and the real forms to encode the parameters of the algorithms. Thus, we can represent agent's plan of roads by the figure 2:

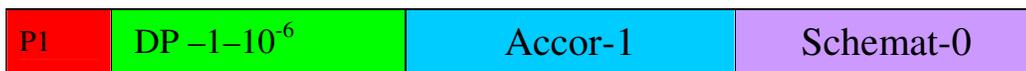

Figure 2. Example of road's plan

### 4.2.2 Plan of Building gene

The agent Building can be characterized by its identifier and by its generalization's algorithms: simplification algorithm and displacement one... etc. The same type of coding is used to encode this type of gene. So, the following figure represents an example of building's plan:

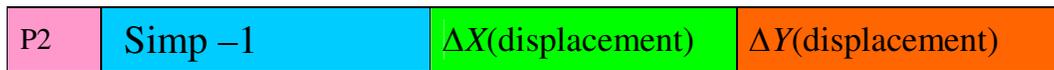

Figure 3. Example of building's Plan

## 4.3 Genetic agent Modules

The main role of a genetic agent is to generalize its self, in order to adapt it to the level of detail requested by the user. For this reason, the genetic agent is responsible for the satisfaction of its constraints. It must be collaborate with the other agent for avoid a constraints violation. It applies the proper solution composed of a sequence of generalization's algorithms, which is generated by its optimizer. The architecture of genetic agent is composed of two main modules. The one is called generalization module which perform the map generalization process and the other is the optimization control module which controls the data flow and the condition of optimization, such as the number of agent in conflicts and the elapsed time of generalization process.

### 4.3.1 Map Generalization module

This module carries out the map's generalization process; it applies the solution defined by the optimizer. The optimizer execute genetic algorithm to define the best chromosome. It is represented by sequence of algorithms and their good parameters, depending on the environment condition and the genetic agent state. For finding the solution, the optimizer follows the classical steps of a genetic algorithm are selection, crossover and mutation [4]. The figure 4 represents a solution generated by the optimizer, such as, each two chromosomes' parents together to make two chromosomes' child, in result. The solution is refined gradually over the iterations until convergence to a solution that approaches the optimal solution, a certain degree of imperfection is acceptable.

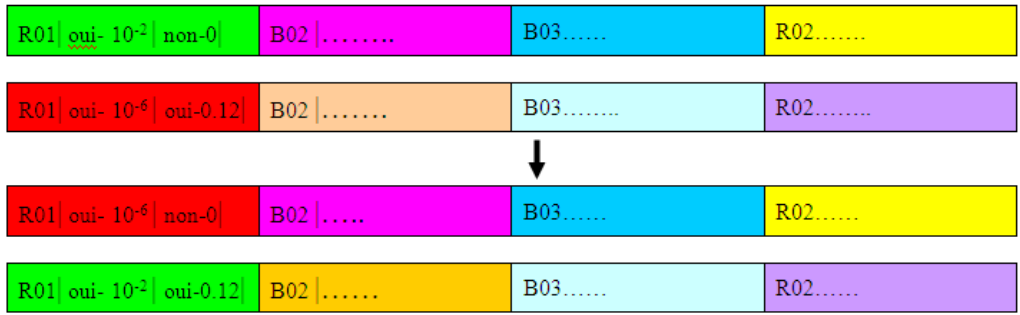

Figure 4. Solution generated by the optimizer

The solution evaluated by a fitness, we use in this work a fitness based on certain measures that assess the quality of spatial data, is restricted to a set of measures:

- OS (object shape): calculates the loss of the object's shape during processing. For buildings, we compute the differentiation in the surface S and for roads, using measures of McMaster [9]:

$$OS = \sum \Delta S + \Delta (\text{McMaster}) \quad (1)$$

   A minimum number is a good solution.

- NC: number of objects in conflict, a minimum number is a good solution.

- DP: sums the normalised, absolute, distance of each object has been displaced from its starting position.

$$DB = \sum_{1}^{n} \sqrt{(dx_i^2 + dy_i^2)} \quad (2)$$

 Also, a minimum number is a good solution.

So, the general function is:

$$f = NC + DP + OS \quad (3)$$

The best solution is one that has the smallest value of general function. After each iteration (selection, crossover, mutation), agents in conflicts exchange the messages for calculate the value of general fitness which allow to select the bests chromosomes, i.e. chromosomes with minimum fitness. After this step, it retains only the most relevant solutions.

### 4.3.2 Optimization Control module

This module controls principally the time constraint and the quality of the map. It controls generalization's time for not exceeds the maximum limits and the number of spatial conflicts. The optimizer stops the execution of the genetic algorithm for three reasons, when it achieve certain fitness, when it achieve a set number of iterations or when it have passed a certain running time. In the two latter cases, we choose the solution that has the best fitness. The optimization control module assures a compromise between the time of delivering spatial data and its quality.

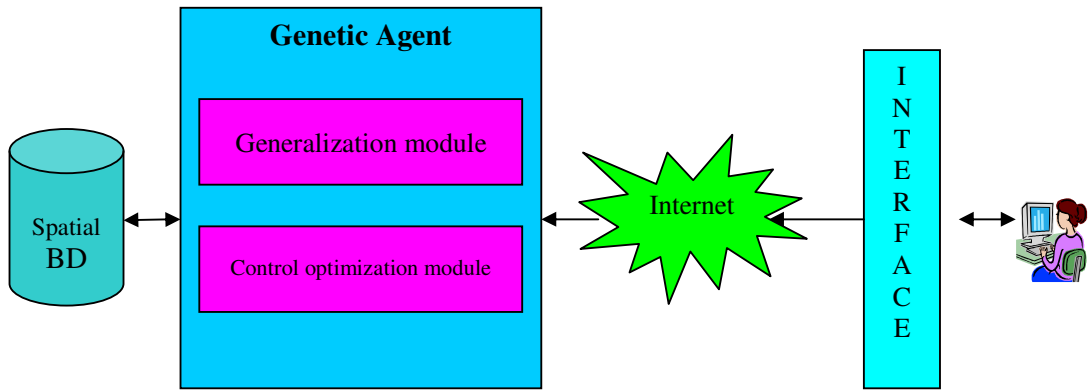

Figure 5.  System architecture

It exists also an interface between the user and the map generalization module; it allows user to transmit its requests. The request transmitted carries important information for the researched data, such as identification of the zone, the kind of map and its level of detail. The level of detail is arbitrary and not predefined. This module permits also to drawing the final generalized data.

In this approach, genetic agents negotiate with each other agent, via message that are used as input in genetic algorithm, these data allow agent to solve various conflicts at once and prevent new conflicts from appearing and achieve a best generalization process.

### 4.4 Cycle life of Genetic Agent

In the beginning of map generalization process, the geographic agent has evaluated its internal constraints and relationalship ones, in its agents group. In our multi-agent system, the agents aim to satisfy the offensive constraints without violating one of defensive constraints [10]. Fitness is utilised for measuring the satisfaction of the constraints (calculate the degree of constraints satisfaction). A list of possible plans is attached which represent the populations of the genetic algorithm. A plan proposes a sequence of generalization's algorithms and its required parameters which improves the agent's state.  The agents active and run theirs genetic algorithms in parallel for reducing the time of map generalization process.

The process of improving an agent's state represented in the following sequence: constraints evaluation, proposing plans which represent the population of genetic, the execution of the genetic algorithm by the optimizer. The generalization process is controlled by the optimization control module for selecting the best plan, the previously steps are the same generic behaviour of all genetic agents, the figure 5 represent the life cycle of agent.

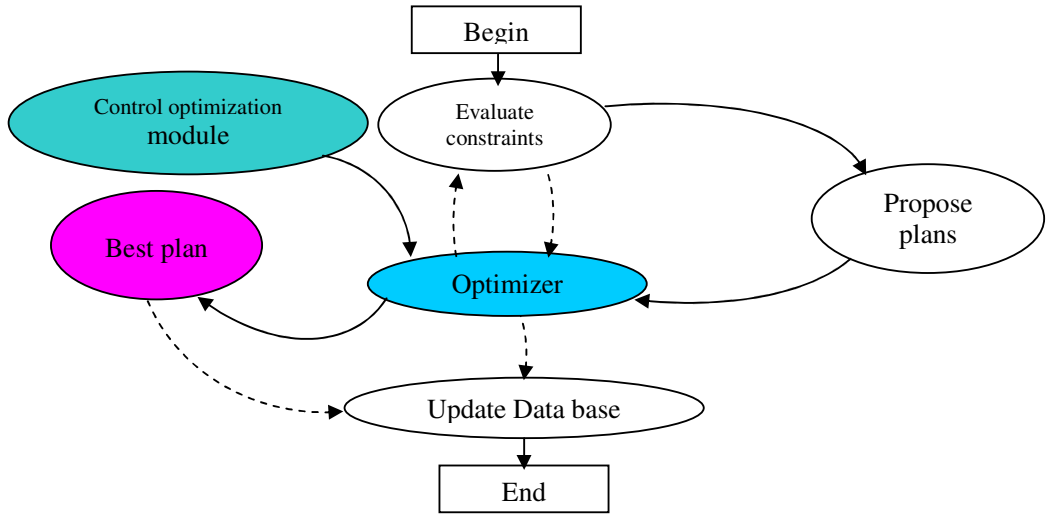

Figure 6. Genetic agent cycle life

## 5. IMPLEMENTATION AND RESULTS

We use in this work the Jade platform [11], JADE is a software framework, fully implemented in Java that simplifies the implementation of multi-agent systems through a middleware. JADE implements FIPA's (Foundation of Intelligent Physical Agents) specifications [2].

The algorithms of generalization and the genetic algorithm are implemented in Java. The agents execute their genetic algorithms in parallel to decrease the response time of the map generalization process. We implement also the various steps of the genetic algorithm; selection, crossover, mutation, and the fitness function for evaluate the solution that was composed of a sequence of generalization's algorithms with good parametric values to improve the map generalization process. Initially, for avoid the complexity; we have used a limited number of objects. We used data in a Shapefile format (ESRI) from the Habitats' direction of el-oued (Algeria). These data are on a scale of 1:1000 and cover a little part of Choot city. Two classes of objects, namely "buildings" and "roads", were used. The scale requested by the user is 1:1500. For drawing the generalized object we use a visualization tool which is based upon the GeoTools library [12]. We present in figure 7 the initial results of our approach.

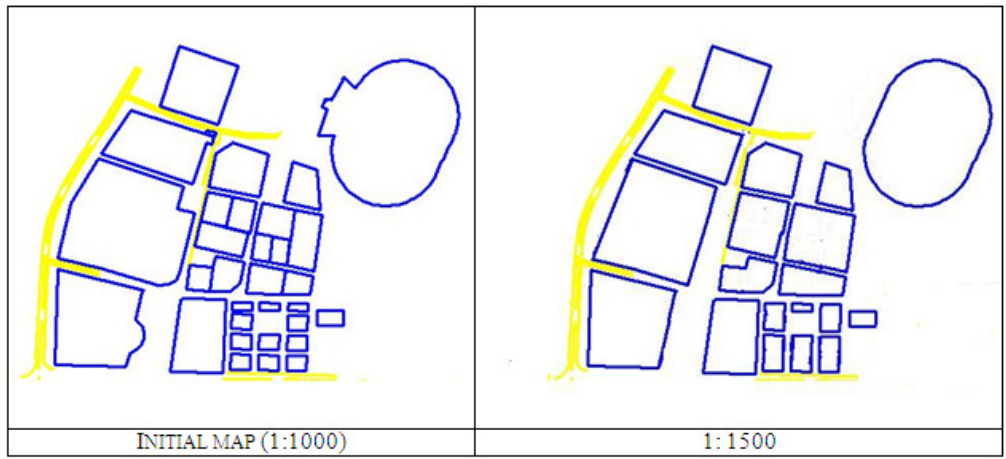

Figure 7. Initial results

## 6. CONCLUSION

In this approach, each type of map features (buildings, roads) is modelled as agent; these agents are equipped with patrimony genetic which allows improving the map generalization process. Multi-agent systems and genetic algorithms provide various advantages to real-time map generalization and quality of generalized data such as, the autonomy and flexibility which improve the personalization of maps at the time of their creations, finding the best solution and resolve all the appearance spatial conflicts. Moreover, the genetic agent has equipped with sensor to check its environment, a communicator to send requests or responses, and the ability to move itself into a new position. Also, to decrease the time of on-the-fly map generalization process, the agents execute theirs genetic algorithms, in parallel and at the same time of the transfer of generalized data.

The initial experiments have shown us the potential advantages of using collaborative agent. Each agent is equipped with patrimony genetic that enable it to generate maps in real-time and with height quality. Thus, each agent can:

- Define the optimal or near optimal actions (sequence of algorithms) of generalization and it can generate its self.

- Adapt its generalization with the other geographic agents.

- Collaborate with the others agent for accelerating the time of map generalization process and resolves all the spatial conflicts.

This research work opens various directions of researches, such as:

- Improving the performance of our approach with the utilization of other methods of optimization that allow the agent to define the optimal solution.

- Customizing the map according to user's needs. In another work, we will model the spatial constraints in the form of ontology for different kinds of map. For example, one can have ontology for constraints semantic of tourist maps. Thus, when using the system, the most appropriate ontology can be selected according to the kind of map that will be generated.

## REFERENCES

[1]     Nafaa jabeur (2006), A multi-agent system for on-the-fly web map generation and spatial conflict resolution, University of Laval, Quebec.

[2]     Mamane Nouri SABO (2007), Généralisation et des patrons géométriques pour la création des objets auto-generalisants (sgo) afin d'améliorer la généralisation cartographique à la volée, Université de laval , Quebec.

[3]     Wilson, I.D. & Ware, M. (1995), Reducing Graphic Conflict In Scale Reduced Maps Using A Genetic Algorithm, ICA Map Generalization Workshop, Paris, 2003.

[4]     Lejdel brahim & kazar okba (2012), Genetic agent for optimizing generalization process of spatial data, International Journal of Digital Information and Wireless Communications (IJDIWC), Vol. 1, No. 3, pp-729-737.

[5]     Van Oosterom & V. Schenkelaars. (1995), The development of an interactive multi-scale GIS, International Journal of Geographical Information Systems, Vol. 9(5):489-507.

[6]     Cecconi, A. (2003), Integration of cartographic generalization and multi-scale databases for enhanced web mapping, PhD. Thesis, University of Zurich.

[7]     A. Newell. (1990), Unified Theories of Cognition, Harvard University Press, Cambridge MA, p549.

[8]     Lejdel brahim & Mrs K. amieur & Zaia alimazighi, (2009), modelling and managing multiples representations of spatial data by a hybrid approach (application them: road network), proceeding of International conference on information & communication systems, pp219-226.

[9]     McMaster, R.B. (1983), Mathematical measures for the evaluation of simplified lines on maps, PhD thesis, University of Kansas., USA.

[10]    Bader M. (2001), Energy Minimizing Methods for Feature Displacement in Map Generalization, PhD thesis, Department of Geography, University of Zurich.

[11]    JADE (2006), Java Agent Development Framework, http://jade.tilab.com/.

[12]    GeoTools (2006), the open source java GIS toolkit, http://docs.geotools.org/.